\input{aipcheck}

\documentclass[
    ,final            
  ]
  {aipproc}

\layoutstyle{6x9}

\def\msun{M$_\odot$}

\begin{document}

\title{Dynamos, Super-pulsars and Gamma-ray bursts}

\author{Stephan Rosswog}{
  address={International University Bremen, Germany}
}

\author{Enrico Ramirez-Ruiz}{
  address={Institute for Advanced Study, Princeton, USA}
}

\begin{abstract}

The remnant of a neutron star binary coalescence is expected to be temporarily
stabilised against gravitational collapse by its differential rotation.
We explore the possibility of dynamo activity in this remnant and assess
the potential for powering a short-duration gamma-ray burst (GRB). 
We analyse our
three-dimensional hydrodynamic simulations of neutron star mergers
with respect to the flow pattern inside the remnant. If the central, newly formed
super-massive neutron star remains stable for a good fraction of a second an 
efficient low-Rossby number $\alpha-\Omega$-dynamo  will amplify the initial 
seed magnetic fields exponentially. We expect that values close to 
equipartition field strength will be reached within several tens of 
milliseconds. Such a super-pulsar could power a GRB via a relativistic 
wind, with an associated spin-down time scale close to the typical 
duration of a short GRB. Similar mechanisms are expected 
to be operational in the surrounding torus formed from neutron star debris.

\end{abstract}

\maketitle


\section{Introduction}

While there is mounting evidence that the long-soft variety GRBs are 
related directly to the death of massive stars and go along with 
supernova explosions, there is so far little evidence about the progenitor
of short-hard GRBs. The most popular candidates are binary coalescences
of either a double neutron star or a stellar mass black hole with a neutron 
star. Most often a 'unified picture' for the GRB central engine, a new-born
black hole plus a debris disk, is invoked. We will explore here the possibility
that the central object of the merger remnant produces a GRB via magnetic 
processes {\em before} collapsing to a black hole. 
Further scenarios with ultra-magnetised neutron stars have been suggested,
for example, by Usov (1992, 1994), Duncan and Thompson (1992), Thompson and 
Duncan (1994), Meszaroz and Rees (1997), Katz (1997) and Kluzniak and Ruderman (1998).\\
The merger of two neutron stars results in a massive central object, 
a thick, hot and dense torus of neutron star debris and some material on
highly eccentric/unbound orbits \cite{ruffert96,ruffert97,rosswog99}.
The central object of the remnant is rapidly differentially rotating 
\cite{rasio99,rosswog99,faber01,rosswog02a} with rotational periods ranging
from $\sim 0.4$ to $\sim 2$ ms \cite{rosswog02a}.
Differential rotation is known to be very efficient in stabilising stars
that are substantially more massive than their non-rotating maximum mass.
For example, Ostriker and Bodenheimer (1968) constructed differentially
rotating white dwarfs of 4.1 \msun. A recent investigation analysing 
differentially rotating polytropic neutron stars \cite{lyford02} finds 
it possible to stabilise systems even beyond twice the typical neutron star
mass of $ 2.8$ \msun. The exact time scale of this stabilisation is 
difficult to determine, as all the poorly known high-density nuclear physics
(``exotic'' condensates etc.) could influence the results, but estimates of 
up to many seconds are not unrealistic.

\section{Simulations}
We have performed 3D simulations of the last inspiral stages and the subsequent
coalescence for about 20 ms. We use a temperature and composition dependent 
nuclear equation of state that covers the whole relevant parameter space in 
density, temperature
and composition \cite{shen98,rosswog02a}. In addition, a detailed,
multi-flavour neutrino treatment has been applied to account for energy losses
and compositional changes due to neutrino processes. The neutrino treatment
and the results concerning the neutrino emission have been described in detail
in \cite{rosswog03a}.
To solve the hydrodynamic equations we use the smoothed particle hydrodynamics 
method (SPH), the simulations are performed with up to more 
than a million SPH particles. We use Newtonian self-gravity plus extra forces
emerging from the emission of gravitational waves. The details of the production 
runs as well as those of several test runs can be found in 
\cite{rosswog02a,rosswog03a,rosswog03b}. Results focusing particularly on 
gamma-ray bursts have been presented in \cite{rosswog02b,rosswog03b,rosswog03c}.

\section{Dynamo action in merger remnants}
\begin{figure}
  \includegraphics[height=.4\textheight]{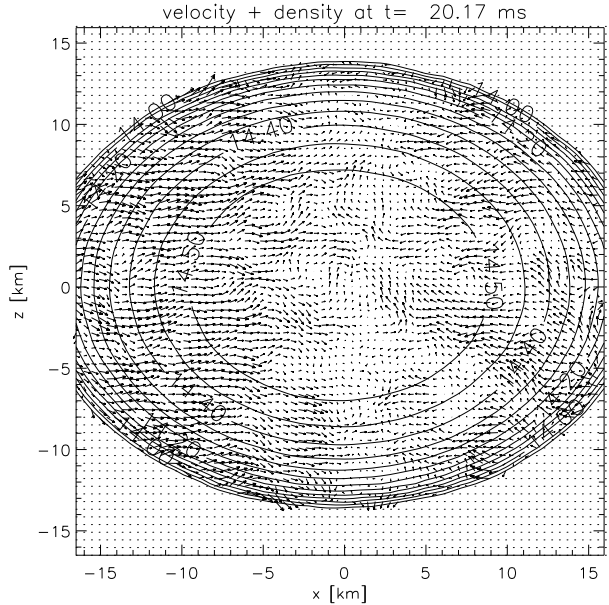}
  \caption{Velocity field (space-fixed frame) inside the central object of
the remnant of a neutron star coalescence. The labels at the contour lines 
refer to log($\rho$), typical fluid velocities are $\sim 10^8$ cm/s.}
\label{vel}
\end{figure}
Before proceeding further with the argumentation, it is worth pointing out that
the fluid flow never becomes axisymmetric during the simulation and 
therefore Cowlings anti-dynamo theorem does not apply here.\\
The central object of the merger remnant is differentially rapidly rotating
with rotation periods below 1 ms over a large fraction of the central object's radius,
examples of rotation profiles can be found in \cite{rosswog02a}. 
When the stellar surfaces come into contact, a vortex sheet forms between them 
across which the tangential velocities exhibit a discontinuity. This vortex sheet is
Kelvin-Helmholtz-unstable with the shortest modes growing fastest. These fluid
instabilities lead complex flow patterns inside the central object of
the merger remnant. In the orbital plane they manifest themselves as strings
of vortex rolls that may merge (see Fig. 8 in \cite{rosswog02a}). An example 
of the flow pattern perpendicular to the orbital plane is shown in  Fig. \ref{vel}.
This pattern caused by fluid instabilities exhibits ``cells'' of size $l_c \sim$ 1 km
and velocities of $v_c \sim 10^8$ cm/s.\\
 Moreover, the neutrino optical depth drops very steeply
from $\sim 10^4$ at the centre to the edge of the central object (see Fig. 11 
in \cite{rosswog03a}). For this reason the outer layers loose neutrino energy, entropy and 
lepton number at a much higher rate than the interior, this leads to a
gradual build-up of a negative entropy and lepton number gradient which will
drive vigorous convection \cite{epstein79,burrows88}. We expect this to set in
after a substantial fraction of the neutrino cooling time (i.e. on time scales longer than
our simulated time) when a lot of the thermal
energy of the remnant has been radiated away. The situation is comparable
to the convection in a new-born protoneutron star, but here we have around twice
the mass and the matter is much more deleptonized ($Y_e \sim 0.1$). As both the fluid
instabilities and the neutrino-driven convection have very similar properties, 
we will not further distinguish between them in this context. Assuming 
neutrinos to be the dominant source of viscosity \cite{thompson93} we estimate 
a viscous damping time scale of the order $\tau_c \sim l_c^2/\nu_{\nu} \sim 60$ s, 
where $\nu_{\nu}$ is the neutrino viscosity. In other words the fluid pattern
will be damped out only on a time scale that is much longer that the time scales 
of interest here.\\
We expect an efficient $\alpha-\Omega$-dynamo to be at work in the merger remnant.
The differential rotation will wind up initial poloidal into a strong toroidal field
(``$\Omega$-effect''), the fluid instabilities/convection will transform toroidal
fields into poloidal ones and vice versa (``$\alpha-$effect''). Usually, the
Rossby number, $Ro \equiv \frac{\tau_{rot}}{\tau_{conv}}$ is adopted as a measure
of the efficiency of dynamo action in a star. In the central object we find Rossby
numbers well below unity, $\sim 0.4$, and therefore expect an efficient amplification of
initial seed magnetic fields. A convective dynamo amplifies initial fields
exponentially with an e-folding time given approximately by the convective 
overturn time, $\tau_c \approx 3 $ ms; the saturation field strength
is thereby independent of the initial seed field (Nordlund et al. 1992).\\ 
Adopting the kinematic dynamo approximation
we find that, if we start with a typical neutron star magnetic field, $B_0= 10^{12}$ G, as seed, 
equipartition field strength in the central object will be reached (provided enough
kinetic energy is available, see below) in only $\approx 40$ ms. The equipartition field
strengths in the remnant are a few times $ 10^{17}$ G for the central object and around
$\sim 10^{15}$ G for the surrounding torus (see Fig. 8 in \cite{rosswog03b}).
To estimate the maximum obtainable magnetic field strength (averaged over the 
central object) we assume that all of the available kinetic energy can be 
transformed into magnetic field energy. Using the kinetic energy stored in the 
rotation of the central object, $E_{\rm kin}= 8\cdot10^{52}$ erg for our generic simulation,
we find $\langle B_{co} \rangle= 
\sqrt{3\cdot E_{\rm kin}/R_{co}^3} \approx 3\cdot 10^{17}$ G 
(note that if only a fraction of 0.1 of the equipartition pressure should be
reached this would still correspond to $\sim 10^{17}$ G).

\section{Gamma-ray Bursts}
There are various ways how this huge field strength could be used to produce a GRB.
The fields in the vortex rolls (see Fig. 8 in \cite{rosswog02a}) will wind up the magnetic 
field fastest. Once they reach field strengths close to the local equipartition value 
they will become buoyant, float up, break through the surface and possibly reconnect in 
an ultra-relativistic blast \cite{kluzniak98}. The time structure imprinted on the 
sequence of such blasts would then reflect the activity of the fluid instabilities inside 
the central object. The expected lightcurve of the GRB would therefore be an erratic
sequence of sub-bursts with variations on millisecond time scales.

Simultaneously such an object can act as a scaled-up ``super-pulsar'' and drive out an 
ultra-relativistic wind. A similar configuration, a millisecond pulsar with a magnetic field
of a few times $10^{15}$ G, formed for example in an accretion-induced collapse, has been
suggested as a GRB-model by Usov (1992, 1994). The kinetic energy from the braking of the 
central object is mainly transformed into magnetic field energy that is frozen in the outflowing
plasma. At some stage the plasma becomes transparent to its own photons producing a blackbody 
component. Further out from the remnant the MHD-approximation breaks down and intense 
electromagnetic waves of the rotation frequency of the central engine are produced. These will transfer
their energy partly into accelerating outflowing particles to Lorentz-factors in excess of $10^6$
that can produce an afterglow via interaction with the external medium.
The other part goes into non-thermal synchro-Compton radiation with typical energies of $\sim 1$ MeV
\cite{usov94}.

\section{Summary}

We have discussed the possibility of dynamo action in the central object created in a 
neutron star merger, which is expected to be stabilised against gravitational collapse 
via differential rotation. If it remains so for a good fraction of a second then the initial neutron
star magnetic fields are expected to be amplified by a low-Rossby number  $\alpha-\Omega$-dynamo. 
In principle enough rotational energy is available 
to attain an average field strength in the central object of $3 \cdot 10^{17}$ G. Locally 
the equipartition field strength (ranging from $10^{16}$ to a few times $10^{17}$ G depending 
on the exact position in the remnant) may be reached. This will cause the corresponding 
fluid parcels to float up and produce via reconnection an erratic sequence of ultra-relativistic 
blasts. In addition the central object can act as a ``super-pulsar'' of $\sim 10^{17}$ G
that transforms most of its rotational energy into an ultra-relativistic wind with frozen-in 
magnetic field. As shown in \cite{usov94} such a wind will result in a black-body component
plus synchro-Compton radiation. Such a super-pulsar will spin-down in $\sim 0.2$ s, just the
typical duration of a short GRB.\\
We have only discussed magnetic processes in the central object of the remnant, but very 
similar processes are expected from the surrounding torus \cite{narayan92}. 
Here, however, longer time scales and lower magnetic field strengths are expected, the 
equipartition fields being around $10^{15}$ G.


\begin{theacknowledgments}
The reported simulations have been performed using the UK
Astrophysical Fluids Facility (UKAFF) and the supercomputer
of the Mathematical Modelling Centre of the University of Leicester
(HEX).
\end{theacknowledgments}


\bibliographystyle{aipproc}   

\bibliography{sample}

\IfFileExists{\jobname.bbl}{}
 {\typeout{}
  \typeout{******************************************}
  \typeout{** Please run "bibtex \jobname" to optain}
  \typeout{** the bibliography and then re-run LaTeX}
  \typeout{** twice to fix the references!}
  \typeout{******************************************}
  \typeout{}
 }

\end{document}